\documentclass[USenglish,twocolumn]{article}

\usepackage[big]{dgruyter}

\usepackage{graphicx}	
\usepackage{amsmath}	
\usepackage{amssymb}	
\usepackage{multirow}
\usepackage{bm}
\usepackage{authblk}

\usepackage{color}
\usepackage{caption}
\setcitestyle{author-year}

\setcitestyle{authoryear,open={(},close={)}}

\newcommand\kms{\ensuremath{\mbox{km}\,\mbox{s}^{-1}}}
\newcommand\Teff{\ensuremath{T_\mathrm{eff}}}
\newcommand\logg{\ensuremath{\log g}}
\newcommand\vsini{\ensuremath{v\sin i}}

\newcommand\met{\ensuremath{[M/H]}}

\newcommand{\aj}{AJ}
\newcommand{\aap}{A \& A}

\newcommand{\mnras}{MNRAS}
\newcommand{\apj}{APJ}
\newcommand{\nat}{Nature}
\newcommand{\rmxaa}{RMXAA}
\newcommand{\apjs}{ApJS}
\newcommand{\aaps}{A \& AS}

\hypersetup{draft}
  
\begin{document}

\author[1]{S. Kassounian$^1$}
\author*[2]{M. Gebran$^1$}
\author[3]{F. Paletou$^{2,3}$}
\author[4]{V. Watson$^{2,3}$}
\runningauthor{S. Kassounian}  

\affil[1]{ Department
of Physics and Astronomy, \LaTeX\ Notre Dame University-Louaize, PO Box 72, Zouk Mika\"{e}l, Lebanon, E-mail: sakokassounian@gmail.com}

\affil[2]{ Department
of Physics and Astronomy, \LaTeX\ Notre Dame University-Louaize, PO Box 72, Zouk Mika\"{e}l, Lebanon, E-mail: mgebran@ndu.edu.lb}

\affil[3]{Universit\'e Paul Sabatier, Observatoire Midi--Pyr\'en\'ees, Cnrs, Cnes, IRAP, F--31400 Toulouse, France, E-mail: frederic.paletou@univ-tlse3.fr}

\affil[3]{\TeX Cnrs, Cnes, Institut de Recherche en Astrophysique et Plan\'etologie, 14 av. E. Belin, F--31400 Toulouse, France.}  

\affil[4]{Universit\'e Paul Sabatier, Observatoire Midi--Pyr\'en\'ees, Cnrs, Cnes, IRAP, F--31400 Toulouse, France, E-mail: viktorwatson@gmail.com}

\affil[4]{\TeX Cnrs, Cnes, Institut de Recherche en Astrophysique et Plan\'etologie, 14 av. E. Belin, F--31400 Toulouse, France.}

\baretabulars 

\articletype{Research Article}

\title{Sliced Inverse Regression: application to fundamental stellar parameters}

\runningtitle{Sliced Inverse Regression}

  \begin{abstract}
{We present a method for deriving stellar fundamental
  parameters. It is based on a regularized sliced inverse
  regression (RSIR). We first tested it on noisy synthetic spectra of A, F, G, and K-type stars, and inverted simultaneously their atmospheric fundamental parameters: \Teff,  \logg, \met\, and \vsini. Different learning databases were calculated using a range of sampling in \Teff, \logg, \vsini, and   \met.  Combined with a principal component analysis (PCA) nearest neighbors (NN) search, the size of the learning database is reduced. A Tikhonov regularization is applied, given the ill-conditioning of SIR. For all spectral types, decreasing the size of the learning database allowed us to reach internal accuracies better than PCA-based NN-search using larger learning databases. For each analyzed parameter, we have reached internal errors that are smaller than the sampling step of the parameter. We have also applied the technique to a sample of observed FGK and A      stars. For a selection of well studied stars, the inverted parameters are in agreement with the ones derived in previous studies. The RSIR inversion technique,
      complemented with PCA pre-processing proves to be efficient in estimating stellar parameters of A, F, G, and K stars.}
\end{abstract}
  \keywords{methods: data analysis, methods: statistical, techniques: spectroscopic,
stars: fundamental parameters.\\}

  \journalname{Open Astronomy}
\DOI{DOI}
  \startpage{1}
  \received{..}
  \revised{..}
  \accepted{..}

  \journalyear{2018}
  \journalvolume{1}

  \maketitle

\section{Introduction}
Astronomical surveys, either spaceborne or ground-based, are gathering
an unprecedented amount of data. One can mention the SDSS DR14 data
\citep{SDSS} that contains 154 TB of millions of spectroscopic and
photometric data. The DR5 of the LAMOST survey \citep{lamost} contains
9 million spectra in total. Gaia DR2
provides information about 1.3 billion stars \citep{gaia}. These space and ground-based surveys quantify the size of the data the astronomical
community will face in a near future.

Spectroscopic analysis is crucial for the derivation of fundamental stellar
  atmospheric parameters which are the
effective temperature (\Teff), the surface gravity (\logg), and the
metallicity (\met). In addition to these fundamentals, and because it
may strongly affect the shape of the observed spectra, the projected
equatorial rotational velocity, \vsini, is also retrieved from
spectroscopic information. Many authors have for long been using
spectroscopic data to estimate the stellar atmospheric parameters
\citep{mcwilliam,latham,torres,correlation,bayesian,HST,Fabro}. However
in order to extract the most relevant and accurate information from
high-resolution, and large bandwidths stellar spectra, still
more endeavour is required.

Most of the traditional approaches and developed pipelines rely on
standard procedures such as comparing an observed spectrum with a set
of theoretical spectra \citep{valenti1996,acess}. The requirement for
advanced computational techniques rises from the generated large
dimensionality of the data due to the wide wavelength coverage
together with high spectral resolution. Many new techniques are
  being developed.
In \cite{Ness} and \cite{cannon2}, a data-driven approach is
introduced (\texttt{CANNON}) for determining stellar labels
(fundamental parameters and detailed stellar abundances) from
spectroscopic data. Their learning databases (LDB) are based on a
subset of reference objects for which the stellar labels are known
with high accuracy.  Dimension reduction techniques are also developed
and used, such as applying the Principal Component Analysis (PCA) for
data reduction (see e.g. \citealt{PCA}). PCA has shown its
effectiveness in inverting the fundamental stellar atmospheric
parameters in several studies \citep{bailer,fiorentin,S4n,
  dms,Gebran}. \cite{kernel} estimated the stellar atmospheric
parameters as well as the absolute magnitudes and $\alpha$-elements
abundances from the LAMOST spectra with a multivariate regression
method based on kernel-based PCA. The LAMOST spectroscopic survey data
has also been recently analyzed by \cite{LAMOST} to invert stellar
parameters and chemical abundances in which several combined
approaches and techniques were compared. The authors developed a code
called \texttt{SP\_Ace} which utilizes nearest neighbor comparison and
non-linear model fitting techniques. In \cite{firefly}, a spectral
fitting code (\texttt{FIREFLY}) was developed to derive the stellar
population properties of stellar systems. \texttt{FIREFLY} uses a
$\chi$-squared minimization fitting procedure that fits stellar
population models to spectroscopic data, following an iterative
best-fitting process controlled by a Bayesian information
criterion. Their approach is efficient to overcome the so-called
``ambiguities'' in the spectra.  More recently, \cite{wavelet} used
wavelet decomposition to distinguish between noise, continuum trends,
and stellar spectral features in the CORALIE FGK-type spectra. By
calculating a subset of wavelet coefficients from the target spectrum
and comparing it to those from a grid of models in a Bayesian
framework, they were able to derive \Teff, \met, and \vsini\ for these
stars. \cite{payne}  presented \texttt{The Payne}, a general method for the precise and simultaneous determination of numerous stellar labels from observed spectra. Using a simple neural-net-like functional form and a suitable choice of training labels, \texttt{The Payne} yields a spectral flux prediction good to $10^{-3}$ rms across a wide range of $T_{\rm eff}$ and $\log g$. \cite{payne}  applied this approach to the APOGEE DR14 data set and obtained precise elemental abundances of 15 chemical species. In the same context, \cite{Fabro} applied a deep neural network architecture to analyse both SDSS-III APOGEE DR13 and synthetic stellar spectra. Their convolutional neural network model, \texttt{StarNet}, was able to predict precise stellar parameters when trained on APOGEE spectra or on synthetic data.\\

In this study, we apply techniques such as, reduction of
  dimensionality with PCA, and a PCA-based nearest neigbor search
\citep{S4n,dms,Gebran} complemented with a Regularized Sliced Inverse
Regression \citep{GRSIR2009,carol} (RSIR) procedure in order to derive
simultaneously \Teff, \logg, \met\, and \vsini\ from spectra of A, and
FGK type stars. Up to now, sliced inverse regression has been rarely
used in astronomy \citep{GRSIR2009,watson}. When combined with 
  PCA-based techniques, the derivations of the fundamental
atmospheric parameters are achieved with higher accuracy compared to
the  sole/mere PCA-based nearest neighbor inversion
\citep{S4n,dms,Gebran}. The mathematical description of our method is
detailed in Sec.~\ref{nums}. Section~\ref{cond} describes all
elements used for the enhancement of the computational abilities of
SIR. Section ~\ref{sims} discusses the application of the technique
on synthetic spectra for A, F, G, and K-type like stars. In
section ~\ref{Obse-stars}, we show the results of inversions of real
stars. Discussion and conclusion are gathered in Sec.~\ref{disc}.

\section{Sliced inverse regression (SIR) }
\label{nums}



SIR, originally formulated by \cite{Li}, is a statistical technique
that reduces multivariate regression to \textit{a lower
  dimension}. It finds an inverse functional relationship between
the response and the predictor which are the fundamental parameters
and the flux respectively. Synthetic spectra flux values, $x_{syn}$,
are usually calculated based on the set of stellar atmospheric
parameters in the form of:
\begin{equation}
x_{syn} = f (\Teff\ , \logg\ , \met\ , \vsini) \, .
\end{equation}

\noindent The inverse functional relation is used to predict the parameters of the observed flux values, $x_{obs}$, in the form of:
\begin{equation}
f^{-1}(x_{obs})=(\Teff\ , \logg\ , \met\ , \vsini) \, .
\end{equation}

In our work, we have derived a functional relationship for each parameter in the following way:

\begin{equation}
Y_j=f^{-1}_j(x_{obs}) \, ,
\end{equation} 
where $j=1,2,3,4$ for \Teff,\,\logg,\,\met, and \vsini.

\subsection{Global covariance matrix $\Sigma$}

SIR starts with the computation of the covariance matrix $\Sigma$ of
all the synthetic spectra $x_i$ of the LDB:

 First, spectra are
gathered in a matrix of dimension $N_{spectra}$ $\times$
$N_{\lambda}$, where $N_{\lambda}$ is
the number of wavelength points per spectrum and
$N_{\mathrm{spectra}}$ is the total number of spectra in the
LDB. Then, the covariance matrix $\Sigma$, is defined as:

\begin{equation}
\label{eq 4}
\Sigma=\frac{1}{N_{\mathrm{spectra}}}\sum_{i=1}^{N_{\mathrm{spectra}}}
( x_i - \overline{x}).( x_i - \overline{x})^T \, ,
\end{equation}
\noindent where the global mean $\overline{x}$ is defined as:
\begin{equation}
\overline{x}=\frac{1}{N_{\mathrm{spectra}}}\sum_{i=1}^{N_{\mathrm{spectra}}}
 x_i \, ,
\end{equation}
$x_i$ being a row vector containing the flux values of spectrum $i$.

\subsection{Intra-slices covariance matrix}

 In SIR, all spectra are organized based on an increasing order of the
  considered parameter for inversion. For example, if we are to
invert \Teff\, of each star, the spectra database should be organized
in increasing order of \Teff\ while having the other parameters ordered
randomly.
We then build-up subsets of spectra, also called ``slices'',
  having the same value of the parameter one wishes to determine
  first. These slices should not overlap each other \citep{Li}.
Then we calculate the means $\overline{x}_h$ of the slice of the
spectra found in each slice $S_h$ that contains $n_h$ synthetic
spectra (h being the index of each slice). For the inversion of each
parameter, $\overline{x}_h$ and $\overline{x}$ are used to calculate
the ``intra-slices'' covariance matrix, $\Gamma$:

\begin{equation}
\label{eq2.9}
\Gamma=\sum_{h=1} ^ H \frac{n_h}{N} ( \overline{x}_h - \overline{x}).( \overline{x}_h - \overline{x})^T  \, ,
\end{equation}
where
\begin{equation}
\overline{x}_h=\frac{1}{n_h}\sum_{x \epsilon S_h} x_i \, .
\end{equation}

\subsection{Dimension reduction and parameter inversion}

SIR aims to build a reducing subspace that \emph{maximizes the variance between the
  slices while minimizing the variance within the slices}  which creates a reduced predictor versus response regressive relationship to predict the parameters of the observed stars. This is applied by the process of stacking the spectra by an increase order of similar or close valued parameters and averaging them into a single spectra and projecting them on a new subspace. These new projection will later be used predictors of the functional relationship. 
Since the reduced projections are formed from spectra  having close parameter values, this insures a higher accuracy of regressive predictions 
\citep{watson}. On the other hand, slicing the spectra based on
non-overlapping similar parameters insures this inter-slice
maximization and intra-slice minimization.
 
The matrix $\Sigma^{-1} \Gamma $ is then calculated where $\Sigma$ and $\Gamma$ are the two previously defined
matrices. One eigenvector of $\Sigma^{-1} \Gamma$, called
$\beta_{\lambda}$ and corresponding to an eigenvalue $\lambda$, is
used to form the reduction subspace. This will allow us to do
regression in a 2-dimensional space using an inverse functional
relationship. This relationship is constructed via a linear piecewise
interpolation between the projection coordinates of the slices on the
single eigenvector of $\Sigma^{-1}\Gamma$ and the parameters.


The selection of $\beta_{\lambda}$, is based on a metric $C_{\lambda}$
that quantifies the relationship between the spectra and the
parameters. $C_{\lambda}$, defined as the ``sliced inverse regression
criteria'' \citep{carol,GRSIR2009}, is calculated as follows:

\begin{equation}
\label{Clambda}
C_{\lambda}= \frac{\beta_{\lambda}^t\, \Gamma\, \beta_{\lambda} }{\beta_{\lambda}^t \, \Sigma \, \beta_{\lambda}}\approx\frac{Var(<\beta_{\lambda}\, .\, x_i >)}{Var(<\beta_{\lambda}\, .\, x_i >)+Var(S_h)} \, ,
\end{equation}
where $\beta_{\lambda}^t$ is the transpose of $\beta_{\lambda}$ and
$Var$ is the variance function

$\beta_{\lambda}^t\, \Gamma\, \beta_{\lambda}$ is the ``inter-slice''
variance, whereas $\beta_{\lambda}^t \, \Sigma \, \beta_{\lambda}$
represents the total variance. The $\beta_{\lambda}$ that gives a
$C_{\lambda}$ value closest to 1 is considered as a the best choice
for the reducing basis vector.
In the present work, $C_{\lambda}$ varies between 0.91 and 0.97 when
using the eigenvector of $\Sigma^{-1}\Gamma$ with the largest
eigenvalue $\lambda$.

\label{sec interpolation}

To invert the parameters, we apply linear piecewise interpolation on
the coordinates of the projections of the $\overline{x}_h$-s on
$\beta_{\lambda}$. Finally, the estimation of the parameters is
made according to:

\begin{equation}
 \widehat{y}= \begin{cases} \overline{y}_1, & \mbox{if } x^{p}\mbox{ $\epsilon ]- \infty,\overline{x}_1^{p}]$ }, \\ \overline{y}_h+\bigg(x_{obs}^{p}-\overline{x}_h^{p}\bigg)\bigg(\frac{\overline{y}_{h+1}-\overline{y}_h}{\overline{x}_{h+1}^{p}-\overline{x}_h^{p}}\bigg) & \mbox{if } x^{p}\mbox{ $\epsilon ]\overline{x}_h^{p},\overline{x}_{h+1}^{p}] $ }, \\ \overline{y}_H, & \mbox{if } x^{p}\mbox{$\epsilon ]\overline{x}^{p}_H,+\infty$[   } , \end{cases}
 \label{piece}
\end{equation}
where $\widehat{y}$ is the estimated parameter; $\overline{y}_h$ is
the mean of the parameters of the spectra in slice $h$. The
superscript \textit{``p''} represents the projected value of a selected
set of data on $\beta_{\lambda}$ i.e., $x^p= <\beta_{\lambda}\, .\, x >$.

\section{Enhancement of the computational abilities of SIR }

\label{cond}
In the present work, we are dealing with large amounts of high
resolution spectra, so that $\Sigma^{-1} \Gamma$ have typical
dimension of $\sim 10^4 \times 10^4$. In addition, using a large LDB
for SIR induces an increase in the intra-slice variance. This will
lead to less accurate inverted parameters. Therefore to simultaneously
address these problems, we applied two additional steps to SIR: first,
using PCA, we reduce the dimension of every spectra in the LDB
\citep{watson} from $\sim 10^4$ to $12$. Second, we apply a
  PCA-based NN-search in the reduced subspace to select a smaller LDB,
  more relevant for the spectra one wishes to analyze. 

 $\Sigma^{-1} \Gamma$ matrix is generally ill-conditioned.
  And the higher the condition number is, the more noise sensitive the
  system becomes \citep{kreyszig}. For the present work, values as
  large as $10^{20}$ were found. In that case $\beta_{\lambda}$ is
  very noise sensitive, leading to an unstable functional
  relationship. As a result, inaccurate inverted parameters may be
  derived.  To solve this issue, we have applied  Tikhonov
  regularization which aims to improve the conditioning of $\Sigma^{-1} \Gamma$ and add a priori information to it based on the analysis of the noise of each observed spectrum. Several regularization methods exist, however, Tikhonov is very common and easily implemented. Other studies may address this issue, such as the truncated SVD used in \cite{watson}. Figure~\ref{fig flowchart} summarizes the successive
  procedures we implemented, and that we discuss with more details
  hereafter.

\subsection{LDB reduction via PCA}
\label{pca}

 PCA is a numerical technique that allows for the reduction of
  dimension of each spectrum by projecting it on a set of orthogonal
  basis vectors called \textit{principal components} (PC's). These
components are the eigenvectors of the global covariance matrix
$\Sigma$.  \cite{S4n} and \cite{Gebran} showed that for databases
similar to the ones used in this study, only 12 PC's associated to the
largest eigenvalues are enough to reduce the LDB, while the
reconstruction error remains less than 1\%. Therefore after this first
pass, the new LDB has dimension of $N_{spectra} \times 12$.

The original
LDB may reach to a dimension of $N_{spectra} \times N_{\lambda}\simeq$
$10^6 \times 10^4$. This is due to the fine sampling in the
parameters, the high dimension of the spectra, and the large
wavelength range which makes the process of SIR computationally heavy
in terms of memory and time.
To reduce the LDB which will be used for SIR, for each observed star, a PCA-based
  nearest neighbor search in the reduced subspace is applied \citep{S4n}. This is done using the ``PCA distance'' $d_j^{(O)}$,
  defined as:

\begin{equation}
 d^{(O)}_j=\sqrt{\Sigma^{12}_{k=1}(\varrho_k - p_{jk})^2}  \, ,
 \end{equation}  
where $\varrho_k$ is the projection coordinate on the $k^{th}$
dimension for an observed spectrum, and $p_{jk}$ is the projection
coefficient on the $k^{th}$ dimension for the $j^{th}$ synthetic
spectrum. Finally for the SIR, a set of NN will be selected for each observed star as we will later describe in sec.~\ref{clustering}. 



\subsection{Tikhonov regularization}
\label{Tikho}

 \begin{center}
 \begin{figure*}
	
   \includegraphics[scale=0.7]{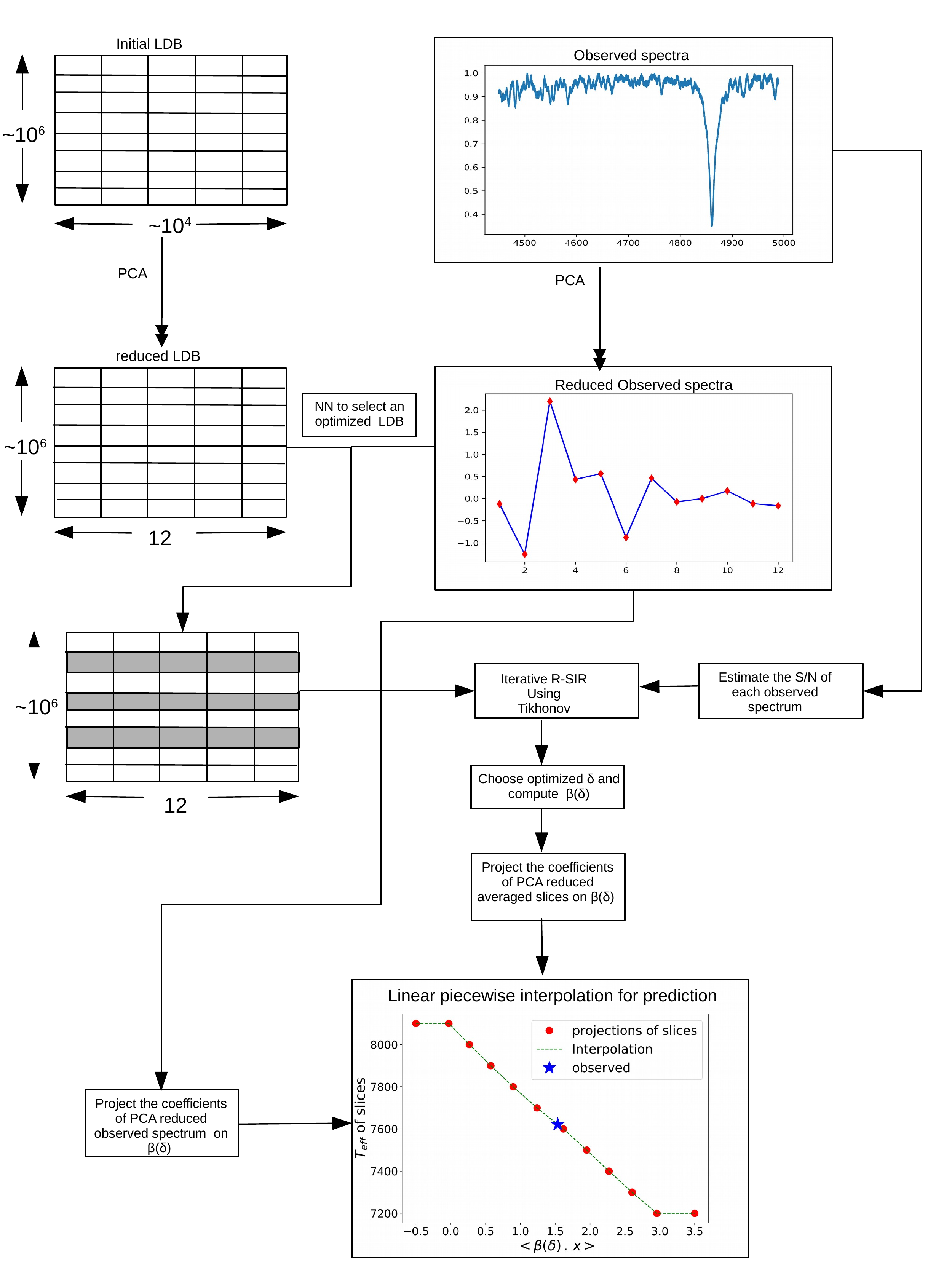}
   
 \caption{ Flowchart of the procedure. The numerics in this figure are for the inversions of the test described in Sec.~\ref{cond}.}
  \label{fig flowchart}
\end{figure*}
\end{center}
 
For the Tikhonov method \citep{tikhonov}, one usually inserts
  a regularization parameter $\delta > 0 $ into the ill-conditioned
  system, usually based on a priori information gathered by analyzing the noise of each observed spectrum. Considering the following matrix:

\begin{equation}
(\Sigma^2 + \delta I)^{-1} \Sigma \Gamma \, .
\label{sigmadelta}
\end{equation}
The eigenvector $\beta_{\lambda}(\delta)$  associated to the
  largest eigenvalue of the matrix defined in Eq.~\ref{sigmadelta} is
calculated based on an optimization approach. For each parameter of
each observed star, an optimum and specific $\delta$ is calculated. This procedure is initiated by estimating the signal to
noise ratio (S/N) of the observed spectrum using the procedure of
\cite{S/N}. Then a random set of synthetic spectra are
    selected from the LDB, and Gaussian white noise having the same
  S/N as the one of the observed spectrum is added to them. SIR is
  finally applied to this selected random set and the prediction of
  their parameters is done via the piecewise interpolation process
  described in Eq.~\ref{piece}. This simulated inversion leads to the selection of an optimum
$\beta_{\lambda}(\delta)$. 

$\delta$ is estimated by minimizing the difference between the newly
inverted parameter values of the randomly selected noise added spectra
($\widehat{y}_i$) and their initial noiseless values ($y_i$). The
comparison is done using a normalized $\chi^2$:

\begin{equation}
\label{nrmse}
\chi^2_N=\sqrt{\frac{\sum_{i=1}^{n}(\,\widehat{y}_i - y_i  )^2}{\sum_{i=1}^{n}(\, y_i-\overline{y}_i )^2}}  \, .
\end{equation}

It was found that $log_{10}(\chi^2_N)$ as a function of $log(\delta)$ is a unimodal function which has a local minimum. This function is displayed in Fig.~\ref{fig log_delta} for a synthetic spectrum having \Teff, \logg, \met\ , \vsini\, and S/N of 7\,600 K, 2.50 dex, 0.0 dex, and 197 \kms, and 196, respectively. The original LDB used in this example is the one of \cite{Gebran}, explained in detail in Sec.~\ref{sims}. To find the minimum of these curves, we applied a golden-section search algorithm \citep{golden}. It is a classical numerical technique that minimizes unimodal functions which have a global minimum. The inversion process for each analyzed spectrum, and each parameter, has its own $\chi^2_N=f(\delta)$ that needs to be minimized.

\subsection{Integrated scheme of the enhancements}
\label{clustering}

Now that we have described the tools that were used to improve the SIR procedure, in what follows we discuss how these techniques are integrated to
increase the accuracy of the inversion process. The flowchart in
Fig.~\ref{fig flowchart} summarizes our adopted approach.

In our work $\Sigma$ and $\Sigma^{-1}
\Gamma$ have reached dimensions of the order of $\sim 10^4 \times 10^4$. For that reason and before applying the SIR process for each spectrum to be analyzed, we have reduced the dimension of these matrices by reducing the size of the original LDB using PCA as described in sec.~\ref{pca}.
 

During SIR, at least two distinct parameter values for each slice are required to construct the functional relationship. Therefore to select the optimum reduced LDB, a test for the construction of this relationship is required. Iteratively we tested for the number of distinct parameters by increasing the number of spectra  of the nearest nearest neighbors. Whenever the values of the distinct concerned parameters become  greater or equal to 2, the iteration breaks and the inversion proceeds to the interpolation.   
During the tests, there were situations where $[\Sigma^2+\delta I]^{-1}$ was singular. This iterative approach solved this problem by adding nearest neighbors. Generally, using a smaller LDB which contain only a set of closest spectra to the observed one theoretically insures the success of SIR compared to using the entire original LDB. When selecting a set of nearest neighbors, we insure a lower minimization value of the intra-slice variance $Var(S_h)$ in Eq.~\ref{Clambda}. Now within each slice the spectra are closer to each other and they are closer to the average spectrum of the slice. At the same time, choosing these optima reduced LDB's overcomes the issue of the degeneracies. In the PCA based NN-search \citep{S4n, Gebran}, we had cases where the $d_j^{(O)}$ where extremely close to each or even equal, with a variety of parameters. In SIR, we do not face such issue because the value are regressed for each parameter and the synthetic spectra with similar or close parameters are averaged to a single slice.     

\begin{center}
 \begin{figure*}[!h]
 \hspace*{-1cm}
   \includegraphics[scale=0.55]{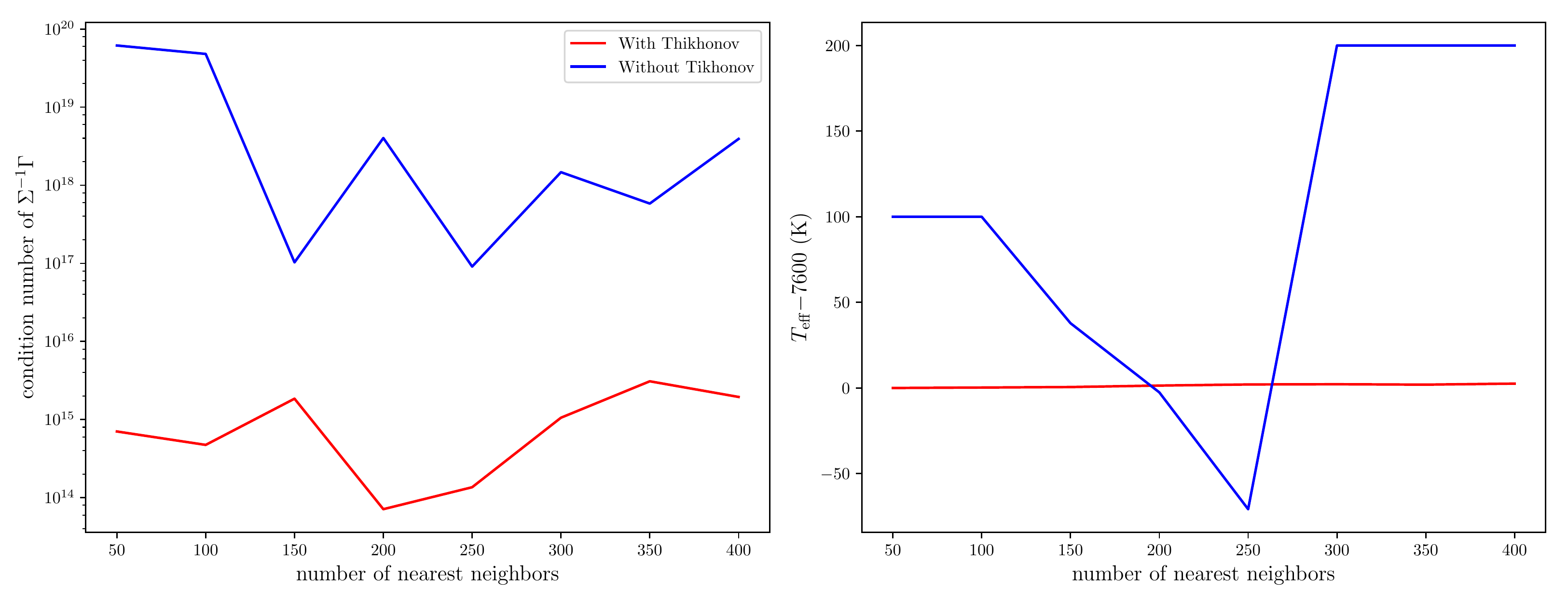}
  \caption{Left: variation of the condition number as a function of the number of nearest neighbor using the optimum reduced LDB with and without the Tikhonov regularization. Right: the inverted \Teff\ as a function of the number of  nearest neighbor with and without Tikhonov regularization. The inversion is done for a noise added synthetic spectrum with  \Teff=7600 K, \logg=2.50 dex, \met=0 dex, \vsini=197 \kms\ and S/N= 196. For clarity we display on the value of \Teff-7600 K  }
  \label{fig Teff_condition_number_syn}
\end{figure*}
\end{center}

Now $\Sigma^{-1} \Gamma$ has a dimension of $12\times12$ and is
constructed from the optimized reduced LDB. Its high condition number
implies that it is ill-conditioned (see the example of the left panel
of Fig~.\ref{fig Teff_condition_number_syn}). Therefore to improve the
inversion process for each observed spectrum, we apply the Tikhonov
regularization in SIR for our selected optima reduced LDB's,
iteratively. By applying this regularization, we are effectively
taking advantage of the S/N ratio analysis and inserting the
propagated noise information as a priori. In other words, we are
applying an denoising procedure.

In Fig.~\ref{fig Teff_condition_number_syn}, we display the inversion
results for \Teff\ of a noisy synthetic spectrum. This spectrum
has a \Teff\ value of 7600 K with an added Gaussian white noise of S/N
= 196. As we iterate over different sizes of optimized reduced LDB's,
a convergence is achieved in every case. For all of our tests, we noticed that the number of spectra in the optimized reduced LDB's did not surpass 500. It is shown in this figure that the condition number of the non-regularized matrix is
$\sim$ 5 orders of magnitude larger than the ones in which the
Tikhonov regularization was applied. The right panel displays the
effect of the regularization on the inverted parameter ( \Teff\ ) of
the same spectrum. It is clearly shown that whatever the number
 of the nearest neighbors in the optimized reduced LDB is, inversion is
  achieved with higher accuracy than the one without
  regularization. The convergence occurs irrespectively of the value of the condition number, as long as it is smaller than the one without Tikhonov  regularization.
  
\begin{figure}[!h]
\centering
\hspace*{-1cm}
   \includegraphics[scale=0.57]{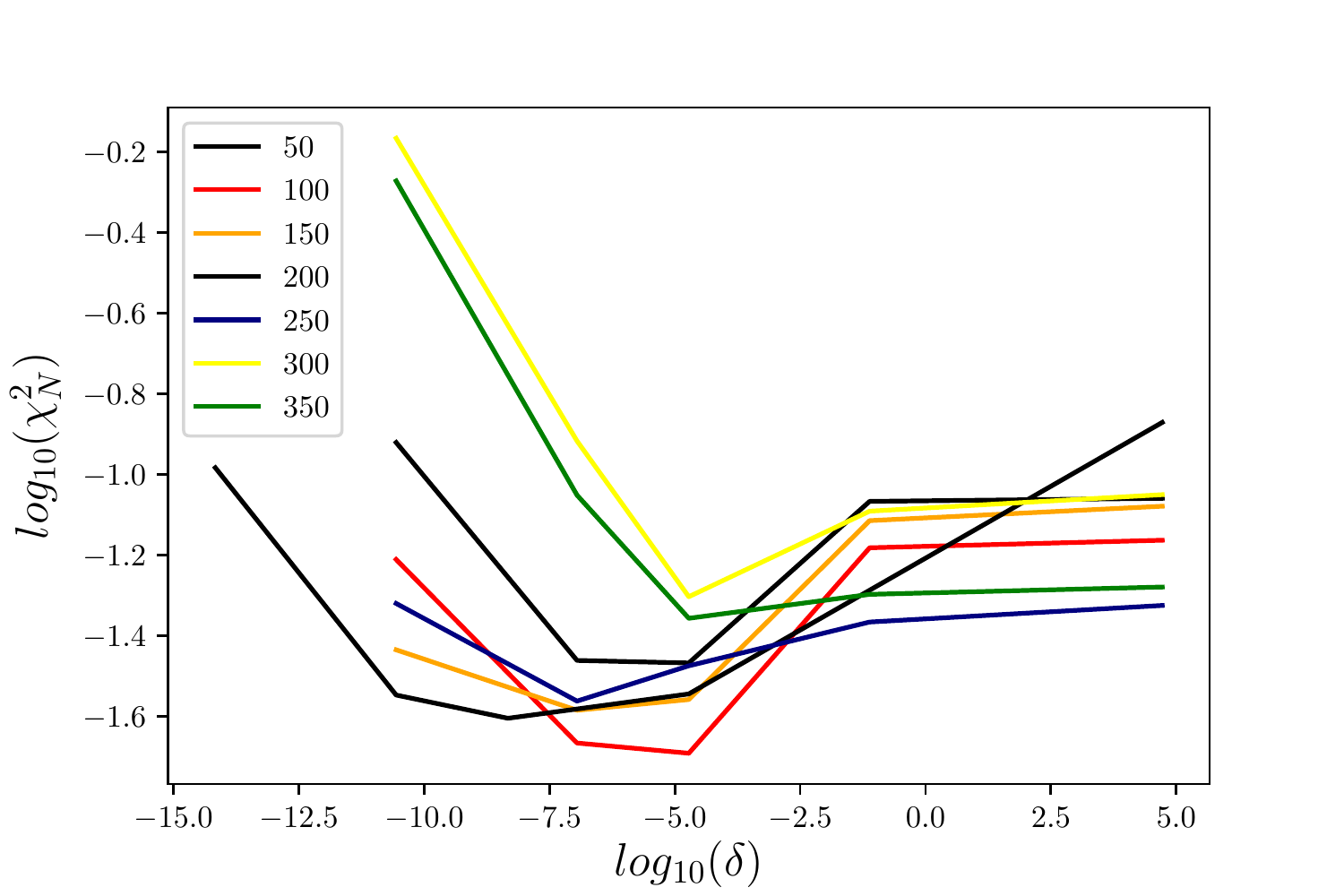}
\caption{$log_{10}(\chi^2_N)$ versus $log(\delta)$ for \Teff\, of a synthetic spectra having \Teff=7\,600 K, \logg=2.50 dex, \vsini=197 \kms, and with a S/N of 196. Each curve displays the minimization using different sets of nearest neigbor generated iteratively.\label{fig log_delta}}
  
\end{figure}

Figure~\ref{fig log_delta} represents the minimization of the
$log_{10}(\chi_N^2)$ as a function of $log_{10}(\delta)$ for different
sets of optimized reduced LDB's. This figure shows the unimodal nature
of the curves irrespective of the size of the LDB.

\section{Simulations and tests}
\label{sims}

In this section, we present the implementation and results of RSIR for two different sets of synthetic spectra. We also compare these results to the ones of the PCA NN-search to show the improvement in the accuracies of the derived parameters.   To each of these spectra, white Gaussian noise was added with a
random S/N. The spectra were calculated in the range of A to K type
stars. The reason for selecting this spectral range is that in
Sec.~\ref{Obse-stars}, we apply this procedure to a sample of the
observed stars studied in \cite{S4n} and \cite{Gebran}.


\subsection{The learning databases}

As done in \cite{dms} and \cite{Gebran},  model atmospheres were calculated using \texttt{ATLAS9}
with the new opacity distribution function
\citep{Kurucz1992,castelli}. These models assume local thermodynamic
equilibrium (LTE), hydrostatic equilibrium, and a 1D plane--parallel
atmosphere. Convection was treated using a mixing length parameter of
0.5 for 7\,000 K $\leq \Teff \leq$ 8\,500 K, and 1.25 for $\Teff
\leq 7\,000$K, following the prescriptions of \cite{smalley2004}.
Synthetic spectra were calculated using \texttt{SYNSPEC48}
\citep{spectra}. The adopted line lists were from Kurucz
gfhyperall.dat\footnote{http://kurucz.harvard.edu} and modified with
more recent and accurate atomic data retrieved from the
VALD\footnote{http://www.astro.uu.se/$\sim$vald/php/vald.php} and the
NIST\footnote{http://physics.nist.gov} databases (for more details see
\citealt{Gebran}). The calculation time for one spectrum depends mainly of the selected wavelength range. For instance, calculating one spectrum in the range of 4450-4990 \AA\ at a resolution of 76\,000 requires $\sim$20 seconds on a personal computer\footnote{Intel core i7-4510U CPU at 2.00GHz $\times$ 4 with 16Gb RAM.}. Around 500 days were necessary for the calculation of the A stars LDB used in Sec.~\ref{SSA}. For the wavelength range between 5000-5400 \AA, one spectrum requires $\sim$22 seconds.

\begin{table}
\centering
\caption{Ranges of the parameters used for the calculation of the A and FGK synthetic spectra LDB's}
\label{table_1}
\begin{tabular}{|c| c|c|} 
 \hline
	Parms & A stars & F/G/K  \\
	\hline

	\Teff\ (K) & [6\,800,11\,000]    & [4\,000,8\,000] \\

	\logg\ (dex)  & $[2.0, 5.0 ]$    & $[3.0, 5.0 ]$ \\

	\met   (dex)  & $[-2.0 , 2.0 ]$  & $[-1.0, 1.0]$ \\ 
	\vsini\ (\kms) & $[0, 300] $     & [0,100]  \\

	$\lambda/\Delta \lambda$ & 76\,000 &50\,000 \\
 	
 	\hline

\end{tabular}

\end{table}

\subsection{Inversion of simulated A stars}
\label{SSA}
  We used the LDB of \cite{Gebran} in which the
 effective temperature of the data varies from 6\,800 up to 11\,000
 K. The wavelength region was chosen between 4\,450$-$4\,990 \AA. This
 wavelength region harbors lines that are sensitive to all stellar
 parameters, and insensitive to microturbulent velocity which was
 adopted to be $\xi_t$=2 km/s based on the work of
 \cite{Gebran,micro}. The adopted resolution is 76\,000 as it
 corresponds to most of the analyzed stars in Sec.~\ref{Obse-stars}. The ranges of all parameters in the A-star LDB are summarized in
 Tab.~\ref{table_1}.

\begin{table*}[hp!]
\center
\caption{Result of inversion for A stars using 3 different LDB's with different steps on 1\,500 noisy synthetic spectra}
\label{table_2}
\begin{tabular}{|c|c|c|c|c|c|c|}

\hline

  Test & Parms & step & $\Lambda_{RSIR}$ & $\Lambda_{PCA}$& offset$_{RSIR}$ & offset$_{PCA}$ \\     	
\hline
  \multirow{6}{*}{ 1} 
& \multicolumn{1}{c|}{\Teff\ (K)}      & \multicolumn{1}{c|}{100}          & \multicolumn{1}{c|}{100.3}   & \multicolumn{1}{c|}{132}   & \multicolumn{1}{c|}{63.45} & \multicolumn{1}{c|}{61.128}     \\[0.2ex] 
& \multicolumn{1}{c|}{\logg\ (dex)}    & \multicolumn{1}{c|}{0.1}          & \multicolumn{1}{c|}{0.12}    & \multicolumn{1}{c|}{0.133} & \multicolumn{1}{c|}{0.042} & \multicolumn{1}{c|}{0.047}     \\  
& \multicolumn{1}{c|}{\met\ (dex)}     & \multicolumn{1}{c|}{0.1}          & \multicolumn{1}{c|}{0.058}   & \multicolumn{1}{c|}{0.066} & \multicolumn{1}{c|}{-0.0102} & \multicolumn{1}{c|}{-0.0057}     \\ [0.1ex] 
& \multicolumn{1}{c|}{\vsini\ (Km/s)}  & \multicolumn{1}{c|}{2: [0-20]}    & \multicolumn{1}{c|}{5.92}    & \multicolumn{1}{c|}{6.47}  & \multicolumn{1}{c|}{-0.279} & \multicolumn{1}{c|}{0.144}     \\ [0.1ex] 
& \multicolumn{1}{c|}{}                & \multicolumn{1}{c|}{5: [20-40]}   & \multicolumn{1}{c|}{}        & \multicolumn{1}{c|}{}      & \multicolumn{1}{c|}{}      & \multicolumn{1}{c|}{}\\ [0.1ex] 
& \multicolumn{1}{c|}{}                & \multicolumn{1}{c|}{10: [40:300]} & \multicolumn{1}{c|}{}        & \multicolumn{1}{c|}{}      & \multicolumn{1}{c|}{}      & \multicolumn{1}{c|}{}\\ [0.1ex]

\hline
      
  \multirow{4}{*}{ 2} 
& \multicolumn{1}{c|}{\Teff\ (K)}      & \multicolumn{1}{c|}{200}    & \multicolumn{1}{c|}{108}    & \multicolumn{1}{c|}{190}   & \multicolumn{1}{c|}{67.54} & \multicolumn{1}{c|}{69.33} \\[0.2ex] 
& \multicolumn{1}{c|}{\logg\ (dex)}    & \multicolumn{1}{c|}{0.2}    & \multicolumn{1}{c|}{0.109}  & \multicolumn{1}{c|}{0.145} & \multicolumn{1}{c|}{0.03} & \multicolumn{1}{c|}{0.04}\\ [0.1ex] 
& \multicolumn{1}{c|}{\met\ (dex)}     & \multicolumn{1}{c|}{0.2}    & \multicolumn{1}{c|}{0.072}  & \multicolumn{1}{c|}{0.107} & \multicolumn{1}{c|}{-0.0102} & \multicolumn{1}{c|}{-0.01}  \\ [0.1ex] 
& \multicolumn{1}{c|}{\vsini\ (Km/s)}  & \multicolumn{1}{c|}{10}     & \multicolumn{1}{c|}{9.63}   & \multicolumn{1}{c|}{10.25} & \multicolumn{1}{c|}{-3.5} & \multicolumn{1}{c|}{-1.82}  \\ [0.1ex] 

\hline
  \multirow{4}{*}{ 3} 
& \multicolumn{1}{c|}{\Teff\ (K)}      & \multicolumn{1}{c|}{400}    & \multicolumn{1}{c|}{174}       & \multicolumn{1}{c|}{295}    & \multicolumn{1}{c|}{40.74} & \multicolumn{1}{c|}{92.39}\\[0.2ex] 
& \multicolumn{1}{c|}{\logg\ (dex)}    & \multicolumn{1}{c|}{0.4}    & \multicolumn{1}{c|}{0.17}      & \multicolumn{1}{c|}{0.224}  & \multicolumn{1}{c|}{0.014} & \multicolumn{1}{c|}{0.061}\\ [0.1ex] 
& \multicolumn{1}{c|}{\met\ (dex)}     & \multicolumn{1}{c|}{0.3}    & \multicolumn{1}{c|}{0.081}     & \multicolumn{1}{c|}{0.113}  & \multicolumn{1}{c|}{-0.016} & \multicolumn{1}{c|}{0.0015} \\ [0.1ex] 
& \multicolumn{1}{c|}{\vsini\ (Km/s)}  & \multicolumn{1}{c|}{20}     & \multicolumn{1}{c|}{12.05}     & \multicolumn{1}{c|}{11.23}  & \multicolumn{1}{c|}{-4.11} & \multicolumn{1}{c|}{-0.9} \\ [0.1ex] 
\hline

\end{tabular}

\end{table*}

Noise added synthetic spectra were calculated to be used as
simulated observations. Around 1\,500 spectra were calculated
for A stars with parameters randomly selected within the range of the LDB but not necessarily at the grid points. To analyze the effect of the sampling on the RSIR technique,
we have inverted these spectra using 3 different LDB's.  
For the same range in all the parameter only the step was modified in each database.  
As an example, in the LDB 1, \Teff\ has a step of 100 K, whereas in LDB's
2 and 3, the steps are 200 K and 400 K, respectively. The same was
done for all parameters and the details about the steps are found in
Tab.~\ref{table_2}. The sampling of
the \vsini\ in the LDB's is not constant and depends on the
value of \vsini\ \citep{Gebran}.

To compare the results of the inversion of PCA NN-search and RSIR for 1\,500 spectra, we estimate the root mean square error for both techniques, $\Lambda$, defined as:

\begin{equation}
\Lambda =\sqrt{ \frac{ \sum_{i=0}^{N} ( \, y_i^{\rm{(inv)}} - y_i^{\rm{(true)}}\, )^2}{N}} \, ,
\end{equation}


where $y_i^{\rm{(true)}}$ is the known parameter of the $i^{th}$ synthetic spectrum and $y_i^{\rm{(inv)}}$ its corresponding inverted one.\\

Columns 4 and 5 of Tab.~\ref{table_2} display the $\Lambda$ results using the PCA NN-search and the RSIR for the 3 LDB's. The offsets, calculated as a signed mean difference, between the inverted and the true values are presented in the last two columns of Tab.~\ref{table_2}. Comparing the $\Lambda$ values of each approach, an improvement is achieved using RSIR for all parameters. One exception exists in the case of \vsini\  for test 3. The large \vsini\ step of the original LDB causes the PCA NN-search pre-processing stage to select inaccurate NN's. 
For most cases RSIR with a coarse sampling in parameters is producing more accurate inversions compared to PCA with a denser sampling. This directly infers a gain in computational time as a coarse sampling leads to smaller LDB. The time required to invert the parameters of one synthetic spectrum depends on the computational facilities. For instance, the gain in time for using the A-stars LDB of test 2 instead of the one of test 1 is $\sim$25\%.

\begin{center}
 \begin{figure*}
   \hspace*{-2cm} 	
   \includegraphics[scale=0.88]{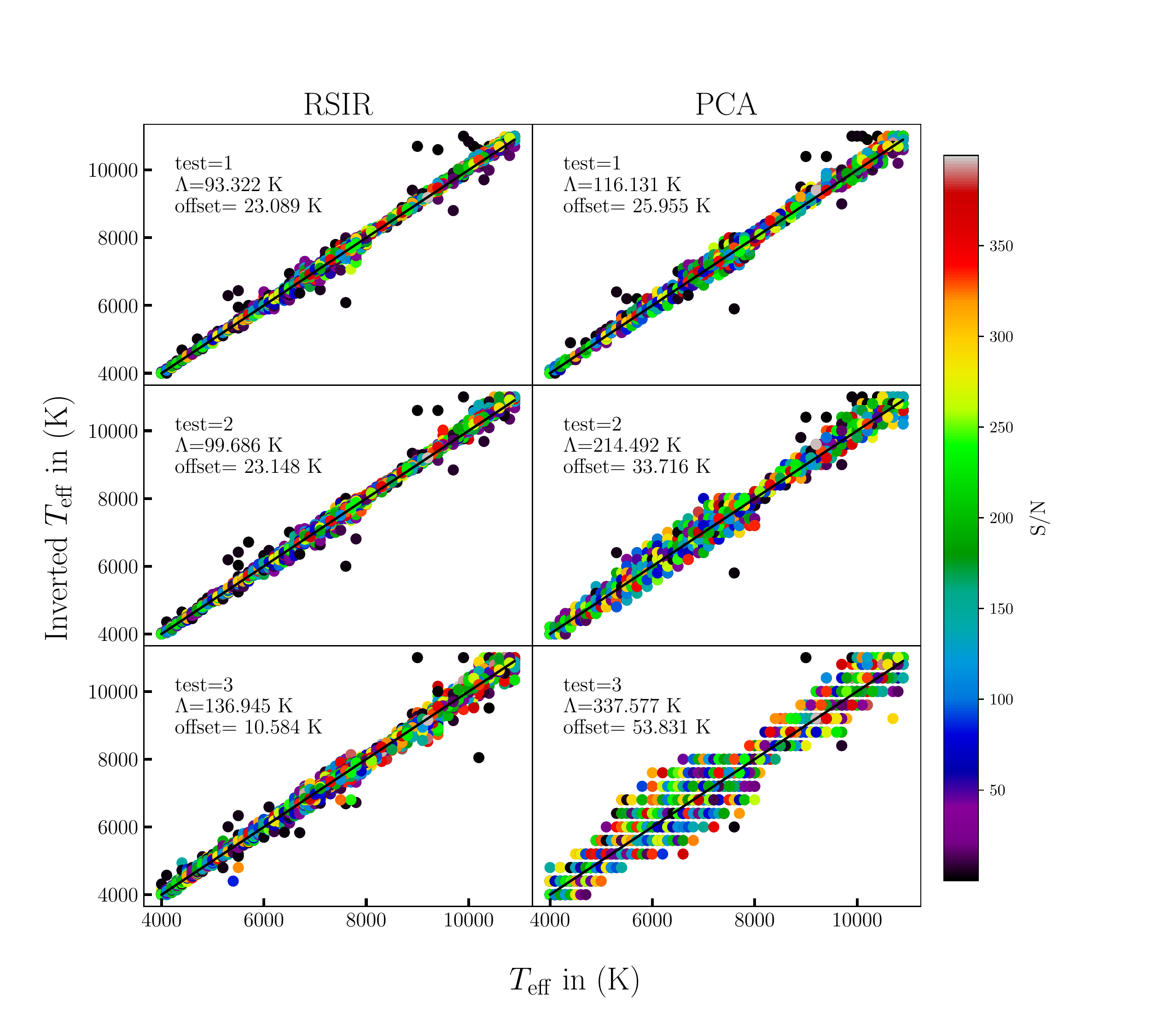}
    \caption{Results of the effective temperature inversion for the synthetic A to K stars. The black line represents the 1-to-1 correspondance.}
  \label{fig Teff_FGK}
\end{figure*}
\end{center}

To analyze the effect of the S/N on the inversions, we display in
Fig.~\ref{fig Teff_FGK} the inverted \Teff\ as a function of the real
\Teff\ for the 1500 A star spectra, for different S/N and
different LDB's (tests 1, 2 and 3). The results of the PCA NN-search is affected both by the sampling size and the S/N of the
analyzed stars, whereas for RSIR, with the pre-processing of PCA and a
Tikhonov regularization this effects becomes less significant on the accuracy of the inversion.  In the appendix, we present the behavior of the inversion of \logg, \met, and \vsini. A similar behavior to the one of \Teff\ can be concluded for these parameters and this can be shown in Figs.~\ref{fig logg_FGK}, \ref{fig meta_FGK} and \ref{fig vrot_FGK}. 

\subsection{Inversion of Simulated FGK type stars}
\label{SSFGK}

The same procedure was applied to FGK star-like spectra. Around 2\,500
noisy spectra were produced in the ranges described Tab.~\ref{table_1}. As done for the A stars, the parameters of the FGK synthetic spectra were also selected randomly. The chosen
resolution of 50\,000 is the same used in \cite{S4n}. The wavelength
range was selected from 5\,000 to 5\,400 \AA\  containing the Mg\,{\sc i} b
triplet, a good indicator of \logg\ and sensitive as well to \Teff. The
microturbulent velocity was set to $\xi_t$ $\sim$ 1 \kms
 \citep{micro}. The inversion results ($\Lambda$ and offsets) as a function of the sampling steps are shown in
Tab.~\ref{table_4}. These results show similar behavior to that of the
A stars in terms of improvement in accuracy while comparing RSIR to PCA NN-search.
In Fig.~\ref{fig Teff_FGK} we also overplot the \Teff\ for our F/G/K noisy synthetic spectra. The effect of inversion as a function of S/N and sampling is very similar to the one of A stars, and for all the parameters (Figs.~\ref{fig logg_FGK}, \ref{fig meta_FGK} and \ref{fig vrot_FGK}).

 \begin{table*}[t!]
\centering
\caption{Result of inversion using 3 different LDB's with different steps on 2\,500 noisy synthetic FGK spectra}\label{table_4}

\begin{tabular}{|c|c|c|c|c|c|c|}

\hline

  Test & parms & step &  $\Lambda_{RSIR}$ & $\Lambda_{PCA}$ & offset$_{RSIR}$ & offset$_{PCA}$  \\     	
\hline
  \multirow{6}{*}{ 1} 
& \multicolumn{1}{c|}{\Teff\ (K)}      & \multicolumn{1}{c|}{100}          & \multicolumn{1}{c|}{106}       & \multicolumn{1}{c|}{117}   & \multicolumn{1}{c|}{5.92} & \multicolumn{1}{c|}{0.04}  \\[0.2ex] 
& \multicolumn{1}{c|}{\logg\ (dex)}    & \multicolumn{1}{c|}{0.1}          & \multicolumn{1}{c|}{0.121}     & \multicolumn{1}{c|}{0.136} & \multicolumn{1}{c|}{0.016} & \multicolumn{1}{c|}{0.002}  \\ [0.1ex]
& \multicolumn{1}{c|}{\met\ (dex)}     & \multicolumn{1}{c|}{0.1}          & \multicolumn{1}{c|}{0.061}     & \multicolumn{1}{c|}{0.063} & \multicolumn{1}{c|}{-0.45} & \multicolumn{1}{c|}{-0.56}  \\ [0.1ex] 
& \multicolumn{1}{c|}{\vsini\ (Km/s)}  & \multicolumn{1}{c|}{2: [0-20]}    & \multicolumn{1}{c|}{1.72}      & \multicolumn{1}{c|}{2.88}  & \multicolumn{1}{c|}{0.005} & \multicolumn{1}{c|}{0.06}  \\ [0.1ex] 
& \multicolumn{1}{c|}{}                & \multicolumn{1}{c|}{5: [20-40]}   & \multicolumn{1}{c|}{}          & \multicolumn{1}{c|}{}      & \multicolumn{1}{c|}{}      & \multicolumn{1}{c|}{}\\ [0.1ex]
& \multicolumn{1}{c|}{}                & \multicolumn{1}{c|}{10: [40-100]} & \multicolumn{1}{c|}{}          & \multicolumn{1}{c|}{}      & \multicolumn{1}{c|}{}      & \multicolumn{1}{c|}{}\\ [0.1ex]

\hline
      
  \multirow{6}{*}{ 2} 
& \multicolumn{1}{c|}{\Teff\ (K)}      & \multicolumn{1}{c|}{200}            & \multicolumn{1}{c|}{109}       & \multicolumn{1}{c|}{236}   & \multicolumn{1}{c|}{8.17} & \multicolumn{1}{c|}{-7.86} \\[0.2ex] 
& \multicolumn{1}{c|}{\logg\ (dex)}    & \multicolumn{1}{c|}{0.2}            & \multicolumn{1}{c|}{0.132}     & \multicolumn{1}{c|}{0.289} & \multicolumn{1}{c|}{0.0027} & \multicolumn{1}{c|}{-0.017} \\ [0.1ex] 
& \multicolumn{1}{c|}{\met\ (dex)}     & \multicolumn{1}{c|}{0.2}            & \multicolumn{1}{c|}{0.066}     & \multicolumn{1}{c|}{0.115} & \multicolumn{1}{c|}{-0.47} & \multicolumn{1}{c|}{-0.62}  \\ [0.1ex] 
& \multicolumn{1}{c|}{\vsini\ (Km/s)}  & \multicolumn{1}{c|}{2: [0-20]}      & \multicolumn{1}{c|}{1.88}      & \multicolumn{1}{c|}{3.17}  & \multicolumn{1}{c|}{0.0087} & \multicolumn{1}{c|}{0.0004} \\ [0.1ex] 
& \multicolumn{1}{c|}{} & \multicolumn{1}{c|}{}                & \multicolumn{1}{c|}{10: [20-40]}    & \multicolumn{1}{c|}{}          & \multicolumn{1}{c|}{}      & \multicolumn{1}{c|}{} \\ [0.1ex] 
& \multicolumn{1}{c|}{} & \multicolumn{1}{c|}{}                & \multicolumn{1}{c|}{10: [40-100]}   & \multicolumn{1}{c|}{}          & \multicolumn{1}{c|}{}      & \multicolumn{1}{c|}{} \\ [0.1ex]

\hline
  \multirow{6}{*}{ 3} 
& \multicolumn{1}{c|}{\Teff\ (K)}      & \multicolumn{1}{c|}{400}          & \multicolumn{1}{c|}{127}       & \multicolumn{1}{c|}{368}   & \multicolumn{1}{c|}{10.43} & \multicolumn{1}{c|}{-27}\\[0.2ex] 
& \multicolumn{1}{c|}{\logg\ (dex)}    & \multicolumn{1}{c|}{0.4}          & \multicolumn{1}{c|}{0.147}     & \multicolumn{1}{c|}{0.45}  & \multicolumn{1}{c|}{-0.01} & \multicolumn{1}{c|}{-0.04}\\ [0.1ex]
& \multicolumn{1}{c|}{\met\ (dex)}     & \multicolumn{1}{c|}{0.4}          & \multicolumn{1}{c|}{0.07}      & \multicolumn{1}{c|}{0.17}  & \multicolumn{1}{c|}{-0.84} & \multicolumn{1}{c|}{-0.625}  \\ [0.1ex]
& \multicolumn{1}{c|}{\vsini\ (Km/s)}  & \multicolumn{1}{c|}{4: [0-20]}    & \multicolumn{1}{c|}{2.25}      & \multicolumn{1}{c|}{5.243} & \multicolumn{1}{c|}{0.018} & \multicolumn{1}{c|}{-0.011}  \\ [0.1ex] 
& \multicolumn{1}{c|}{}                & \multicolumn{1}{c|}{10: [20-40]}  & \multicolumn{1}{c|}{}          & \multicolumn{1}{c|}{}      & \multicolumn{1}{c|}{}      & \multicolumn{1}{c|}{}   \\ [0.1ex]
& \multicolumn{1}{c|}{}                & \multicolumn{1}{c|}{10: [40-100]} & \multicolumn{1}{c|}{}          & \multicolumn{1}{c|}{}      & \multicolumn{1}{c|}{}      & \multicolumn{1}{c|}{}\\ [0.1ex]

\hline

\end{tabular}

\end{table*}

\vskip 1cm
\section{Application to observed spectra}
\label{Obse-stars}

\begin{figure*}[!h]
\centering
	\hspace*{-2cm}
   \includegraphics[scale=0.75]{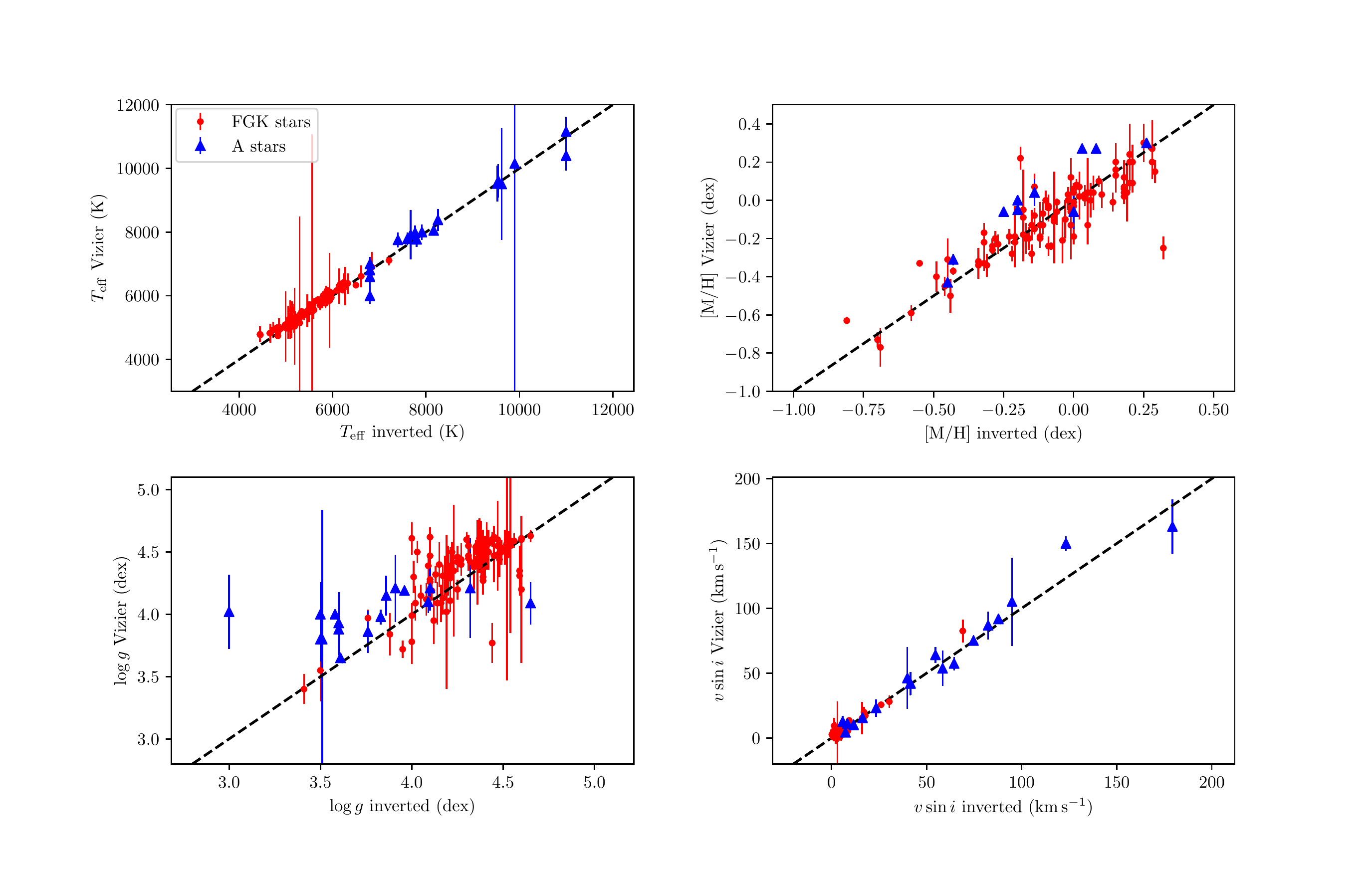}
  \caption{Comparison between the inverted parameters for our FGK stars sample (filled circles) and A stars (filled triangle). }
  \label{fig s4n}
\end{figure*}
The performance of the RSIR has been tested on two samples of stellar
spectra. The first sample is the one of the Spectroscopic Survey of
Stars in the Solar Neighborhood (S$^4$N,
\citealt{2004A&A...420..183A}). These are spectra of bright FGK stars
that are at distance less than 15 pc. We have estimated the S/N of
these spectra in the wavelength range used for the inversion of the
parameters [5\,000$-$5\,400\AA]. This ratio ranges between 40 and
450. These spectra are at a resolution of $\lambda/\Delta \lambda
\sim$50\,000. All the details about the acquisition and the reduction
procedure of the S$^4$N data can be found in
\cite{2004A&A...420..183A}. These spectra were inverted using the
database of \cite{S4n}, made of 905 spectra retrieved from the ELODIE
stellar library \citep{elodie1,elodie2}. We have used the Mg\,{\sc i} b
triplet wavelength range as explained in Sec.~\ref{SSFGK}. We then
compared the values of the inverted parameters with the ones of
\cite{2004A&A...420..183A} and to the medians found in the Vizier
catalog\footnote{The query was performed using the method described in
  \cite{query}} for all these stars. The main reason for using this catalog for our comparison is the necessity for reliable and objective catalogs which are constructed based on previous adopted values by the astronomical community.
  
Comparing our inverted \Teff\ to
the ones of \cite{2004A&A...420..183A}, we found an average signed
difference of 2.09 K with standard deviation of 102 K. For \logg, the
average signed difference and the standard deviation are both 0.15
dex. For \met\ and \vsini, we found -0.06$\pm$0.08 dex and
-0.21$\pm$1.89 \kms, respectively.  If we compare our inverted values
to the median of Vizier, we find -85$\pm$110 K, -0.07$\pm$0.16 dex,
0.01$\pm$0.10 dex and -0.50$\pm$2.25 \kms\ as a signed mean difference
and a standard deviation between the catalogued values and the
inverted ones for \Teff, \logg, \met\ and \vsini, respectively. Figure
\ref{fig s4n} displays in filled circles, for the four parameters, the
comparison between our inverted values for the FGK observed spectra
and the ones derived from Vizier. We have also assigned the catalogues
values an error bar corresponding to the standard deviation of the
dispersion in the catalogues values for each star.\\

 \begin{figure*}
\centering
   \includegraphics[scale=0.5]{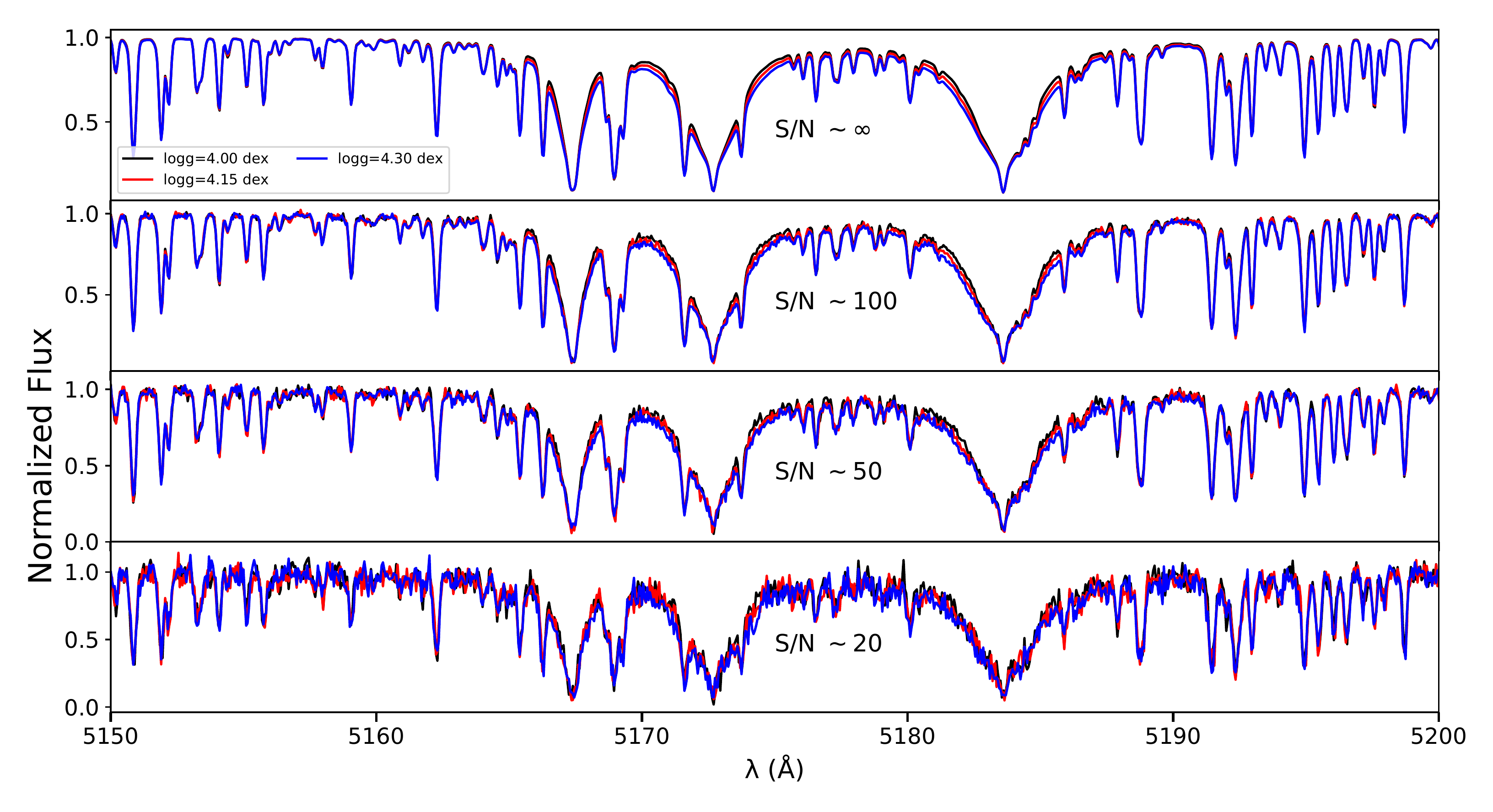}
  \caption{Synthetic spectra of G stars having [\Teff,\logg,\met,\vsini] of 5200 K,[4.00,4.15,4.30 dex], 0.0 dex, 6 \kms. Each plot displays the spectra with different S/N.}
  \label{fig FGKSNR}
\end{figure*}

The second sample of our analysis is constituted of the well studied A
stars of \cite{Gebran}. These are the 19 stars that have been studied
extensively by different authors using different techniques (Vega,
Sirius A, HD~22484, HD~15318, HD~76644, HD~49933, HD~214994,
HD~214923, HD~113139, HD~114330, HD~27819, HD~5448, HD~33256,
HD~29388, HD~91480, HD~30210, HD~32301, HD~28355, and HD~222603) and
have more than 120 references each. The source of these high
resolution spectra is explained in detail in \cite{Gebran}. They were
observed using ELODIE, NARVAL, ESPaDOnS and SOPHIE
spectrographs. ELODIE has a resolution of 42\,000 whereas NARVAL,
ESPaDOnS and SOPHIE are at a resolution of $\sim$76\,000.\\ 
We have
applied the RSIR on these data using the database of Test 1 in
Sec.~\ref{SSA} at both resolutions. The S/N of these spectra is
between 180 and 360.  The inverted parameters of each star were
compared to the ones retrieved from Vizier and added to the plots of
Fig.~\ref{fig s4n} as filled triangles. We found an average signed
difference and a standard deviation of -0.14$\pm$245 K,  -0.20$\pm$0.30
dex, -0.11$\pm$0.09 dex and -2.07$\pm$8.5 \kms\ for \Teff, \logg,
\met\ and \vsini, respectively between the inverted and the Vizier
parameters.

\subsection{The case of \logg}

These results show that most of our inverted parameters are in agreement
with previous studies. Considering that the most accurate parameters
of these A and FGK stars are the Vizier median, our values are less
spread with respect to the median than the ones of \cite{S4n} and
\cite{Gebran}. The standard deviations that we found could be assigned
as an estimation of the errors on the derived parameters. We can
therefore assign precision of 110 K, 0.16 dex, 0.10 dex and 2.25
\kms, on \Teff, \logg, \met, and \vsini, respectively for FGK
stars. For A stars, we found precision of 245 K, 0.30 dex, 0.09 dex,
and 8.50 \kms, on \Teff, \logg, \met, and \vsini, respectively. These
errors are summarized in Tab.\ref{error-obs}.

\begin{table}\small
\centering
\caption{Estimation of the offset (signed mean difference) and the errors on the derived parameters for FGK and A stars.}
\begin{tabular}{|c|c|c|c|c|}

\hline
\baretabulars
Parameter& Offset (FGK) & $\sigma_{\mathrm{FGK}}$ & Offset (A) & $\sigma_{\mathrm{A}}$\\ 
\hline
\Teff\ (K) &-85 & 110 &-0.14& 245 \\ \hline
\logg\ (dex) & -0.07&0.16 & -0.20& 0.30 \\ \hline
\met\ (dex) &0.01& 0.10 &-0.11& 0.09 \\ \hline
\vsini\ (\kms) & -0.5&2.25 &-2.07& 8.50 \\ \hline
\end{tabular}\label{error-obs}

\end{table}

Surface gravity is systematically the most difficult parameter to
determine, with typical errors of the order of 0.15 to 0.3
dex. This parameter is very important for chemical analysis as some line profiles could be very sensitive to $\log g$ values. Spectroscopic determinations of surface gravity have always been
assigned moderately large error bars, especially for A stars
\citep{2005MSAIS...8..130S}. The same applies to FGK stars but with
smaller error bars. Asteroseismic \logg\ determinations remain the
best tools for achieving accuracies less than 0.05 dex
\citep{2013A&A...556A..59H,2013MNRAS.431.2419C,2014ApJS..210....1C}. RSIR
is mainly based on finding the best set of spectra in the database
that correspond to the observed one. As it is a spectroscopic method,
we should not expect an accurate recovery for \logg. Using our values
for $\Delta {\mathrm log}g$ we can, a posteriori figure out what it
means in terms of discernibility between two spectra whose respective
\logg\ differ from this quantity. This also gives us relevant
information about $(i)$ which specific bandwidth(s) are the most
sensitive to such differences, and $(ii)$ how significant they are for
various S/N.

\begin{figure*}
\centering
   \includegraphics[scale=0.5]{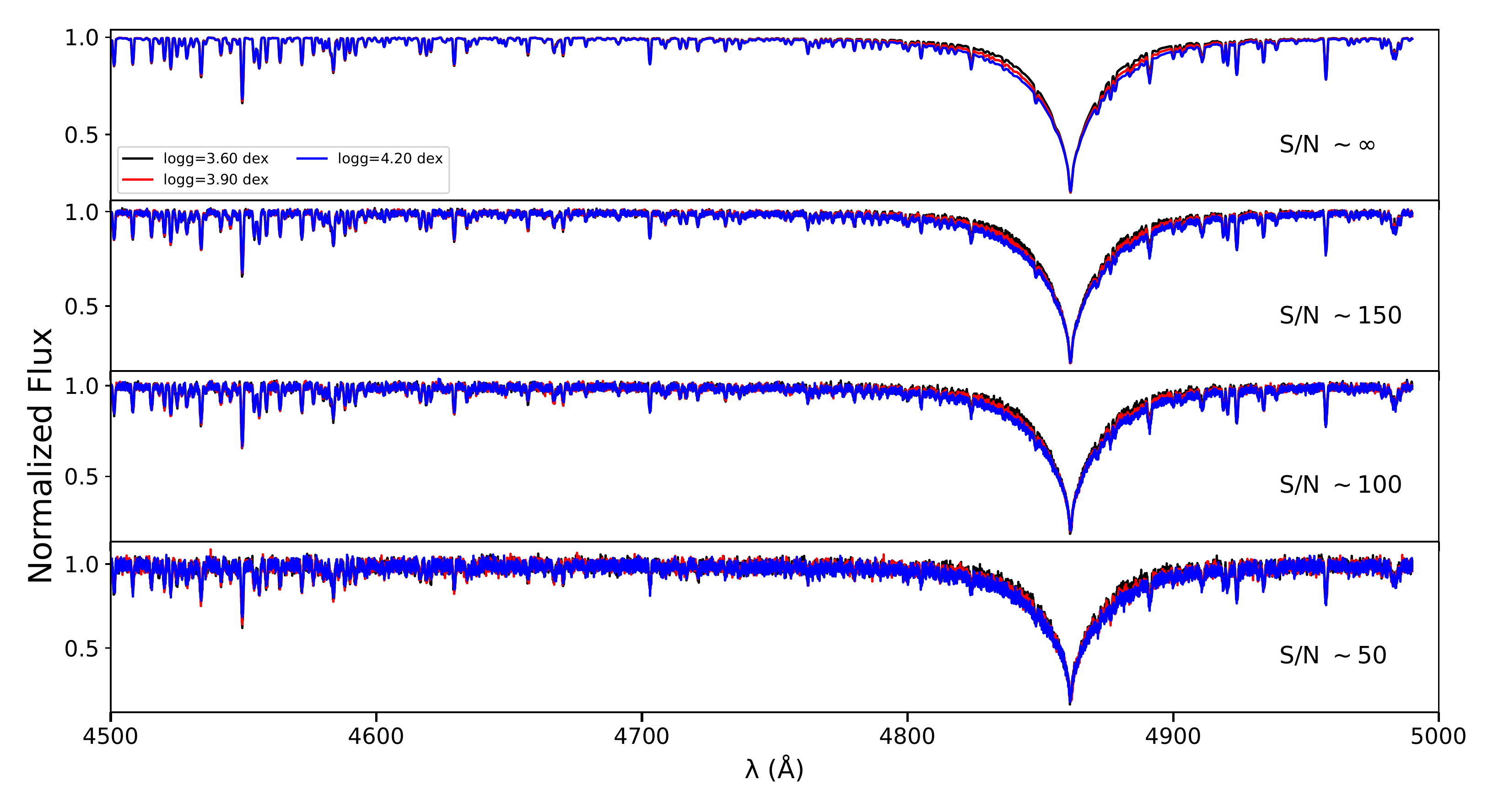}
  \caption{Same as for Fig.~\ref{fig FGKSNR} but for A stars having having [\Teff,\logg,\met,\vsini] of [8\,500 K,[3.60,3.90,4.20 dex], 0.0 dex, 40 \kms.}
  \label{fig ASNR}
\end{figure*}

Figures \ref{fig FGKSNR} and \ref{fig ASNR} display the variation in
the spectrum profile as a function of \logg, fixing all the remaining
parameters, for G and A stars, respectively. In Fig.~\ref{fig
  FGKSNR}, we calculated synthetic spectra for a typical G star
with a \Teff\ of 5\,200 K, \met\ of 0.0 dex, \vsini\ of 6 \kms\ at a
resolution of 50\,000 and in the wavelength range of 5\,000$-$5\,400
\AA. The only parameter that differs between the 3 spectra is \logg,
ranging between 4.00 dex and 4.30 dex with a step of 0.15 dex. The
upper panel of Fig.~\ref{fig FGKSNR} displays the normalized flux of
the synthetic spectra in the Mg\,{\sc i} b triplet region. The flux
level in this region is very sensitive to variation in \logg. The
following panels displays the same spectra for different values of
S/N. When no noise is added (panel with S/N$\sim \infty$), the
distinction between the 3 spectra is clear but when S/N starts to
decrease, the distinction between the noisy spectra becomes harder to
detect. This shows that for a S/N in the order of 100, the noisy
spectra with \logg\ of 4.00 and 4.15 dex are very similar and
therefore the best corresponding synthetic spectrum in our LDB could
have a \logg\ varying at least 0.15 dex from the correct value. We are
not trying to quantify the minimum S/N required for an accurate
inversion of \logg\ as the RSIR is not based on a pixel-to-pixel
comparison, but we are showing the effect of our derived standard
deviations in \logg\ on the flux for noisy spectra.  Figure \ref{fig
  ASNR} displays a similar behaviour for A stars having similar
\Teff\ of 8\,500 K, \met\ of 0.0 dex, \vsini\ of 40 \kms, at a
resolution of 76\,000 in the wavelength range of 4\,500-5\,000
\AA. Surface gravity of these spectra ranges between 3.60 and 4.20 dex
with a step of 0.30 dex. This figure shows a similar behavior to that of Fig.~\ref{fig FGKSNR}. At a S/N of $\sim$150, the
distinction between spectra having a difference of 0.30 dex in \logg,
becomes hardly noticeable. Figures \ref{fig FGKSNR} and \ref{fig ASNR}
also show that the effect of weak metallic lines, on the derivation of
\logg, becomes negligible as the S/N decreases. The \logg\ information
that these lines contain is mainly lost in the noise.

\section{Discussion and conclusion}
\label{disc}

RSIR tests for nearly 4\,000 synthetic stars of different spectral
type and different noise levels showed an improvement over the
PCA-based method of \cite{S4n,dms} and \cite{Gebran} for the inversion
of stellar parameters. Results of Tabs.~\ref{table_2} and
\ref{table_4} and Fig.~\ref{fig Teff_FGK}, for FGK and A stars, show
that for most of the tests, the $\Lambda$ values of RSIR are lower
than nearest neighbor PCA approach. Having a prior information about
the star using PCA as a pre-process allows us to narrow down the
selection of the optimized reduced databases. This decreases drastically the
size of the LDB's. Achieving lower $\Lambda$ with bigger steps helps in
decreasing the prohibitive computation time for the construction of
databases and the calculations of the PC's. Simulated tests revealed that
computation time of RSIR is nearly 1\% of that of the process of PCA
nearest neighbor approach. 

One should be very careful while increasing the size of steps of the
parameters because the PCA pre-processing step could deviate
drastically from the true inverted values therefore excluding the
spectra that actually best describes the observed ones.

Application to observed FGK and A stars reveal a good agreement
between the inverted parameters and the ones derived in previous
studies. The comparison with Vizier catalog values show an improvement
in the derived parameters as compared to the results of \cite{S4n} and
\cite{Gebran} for the same stars and LDB's. Surface gravity remains the
parameter with the least accuracy. Our derived errors on \logg\ are in
the order of 0.15-0.30 dex. Smarter LDB's should be therefore considered,
say, ``adaptive sampling'' (in the parameters under study), taking care
with more caution of the flux typical variations at the most sensitive
wavelength (sub-)domains, \emph{together with} the S/N of the
observations, instead of the a priori sampling in the parameters.
Also, a commonly reported issue with the inversion of stellar
parameters using a LDB of synthetic spectra are the so-called
``ambiguities''. This means that two sets of distinct parameters may
generate spectra which are beyond ``discernibility''. Given a set of
observed spectra to characterize, we could naturally relate that
discernibility to their level of S/N. Using a nearest neighbor search
PCA-based method, for instance, such a level of S/N can easily be
translated into a threshold of distance $\delta_{\rm PCA}$. Then, we
can anticipate that, instead of relying on LDB's usually made using a
priori sampling \textit{in the parameters}, a smarter DB should rely
on $\delta_{\rm PCA}$ instead. This would imply to set up LDB's for fundamental stellar parameters in a radically different fashion
vs. common practices. In the frame of PCA, it would be more relevant
to sample properly the full range of parameters with a
``constrained-random'' process ensuring that there are no nearest
neighbors closer than $\delta_{\rm PCA}$. Such a ``sieve algorithm''
was first proposed by \cite{2002ApJ...575..529L} in the context of the
characterization of magnetic fields from spectropolarimetric data (see
also \citealt{2013ApJ...773..180C}). Another line of development relates to the "structure" of our LDB's. Smarter, or optimal LDB's, using different methods of samplings, should be considered. Such a general issue was already evoked by \cite{Bijaoui2012} for instance. 

Available online databases are usually calculated with large steps in
\Teff\ and \logg. Our RSIR technique, as it does not require small
steps in the LDB, is a good tool to be used with online available
synthetic spectra such as the POLLUX\footnote{pollux.oreme.org}
database \citep{pollux} that contains models with temperature ranging
between 3\,000 and ~50\,000 K or TLUSTY Non-LTE Line-blanketed Model
Atmospheres of O-Type Stars \citep{tlusty} with \Teff\ ranging between
27\,500 and 55\,000 K with 2\,500 K steps, and \logg\ between 3.0 and
4.75 with steps of 0.25 dex. We can also mention the PHOENIX
\citep{phoenix} models database for stars having \Teff$<$12\,000 K and
the AMBRE \citep{ambre} project that contains high-resolution FGKM
stellar synthetic spectra.

As an output of the new Gaia Data Release 2, \cite{RVSdr2} describe the Gaia RVS specification as well as the predicted performance at the end of the mission. Gaia RVS will provide us with a large number of spectra in the calcium
triplet regime (845$-$872 nm). This triplet is very sensitive to
\Teff\ and \logg. The medium resolution (11\,500) of the RVS and the
small range in wavelength would require LDB smaller than the ones used
in our work, leading to a fast application of the RSIR. As we did for
the inversion of the S$^4$N data in Sec.~\ref{Obse-stars}, LDB could
be constructed with real observed stars having well known fundamental
parameters and with the same resolution. Finally, since RSIR is based
on single parameter inversion process, one can also incorporate other
parameters at the cost of computing and handling more numerous individual spectra, for example, microturbulence velocity and individual chemical abundances. 

\nopagebreak

\section*{Appendix}

In this appendix we display the results of the inversion of \logg, \met, and \vsini.  The black line corresponds to the 1-to-1 associated values. The test number, the root mean square error  $\Lambda$, and the offset (see Tabs.~\ref{table_2} and \ref{table_4} for details) are presented.
\begin{center}
  \begin{figure*}
   \hspace*{-2cm} 	
   \includegraphics[scale=0.88]{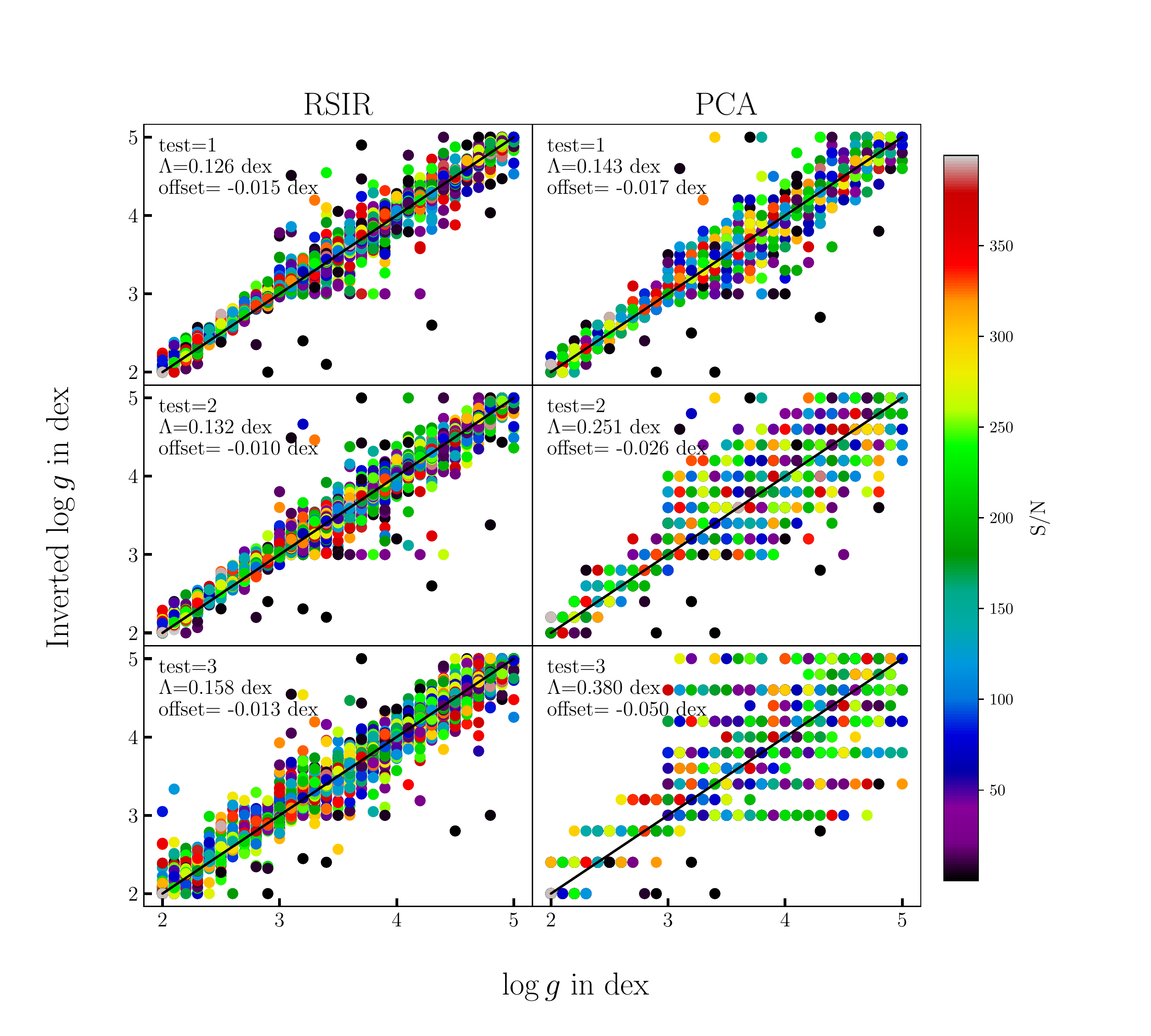}
    \caption{Results of the surface gravity inversion for the synthetic A to K stars. }
  \label{fig logg_FGK}
\end{figure*}
\begin{figure*}
   \hspace*{-2cm} 	
   \includegraphics[scale=0.88]{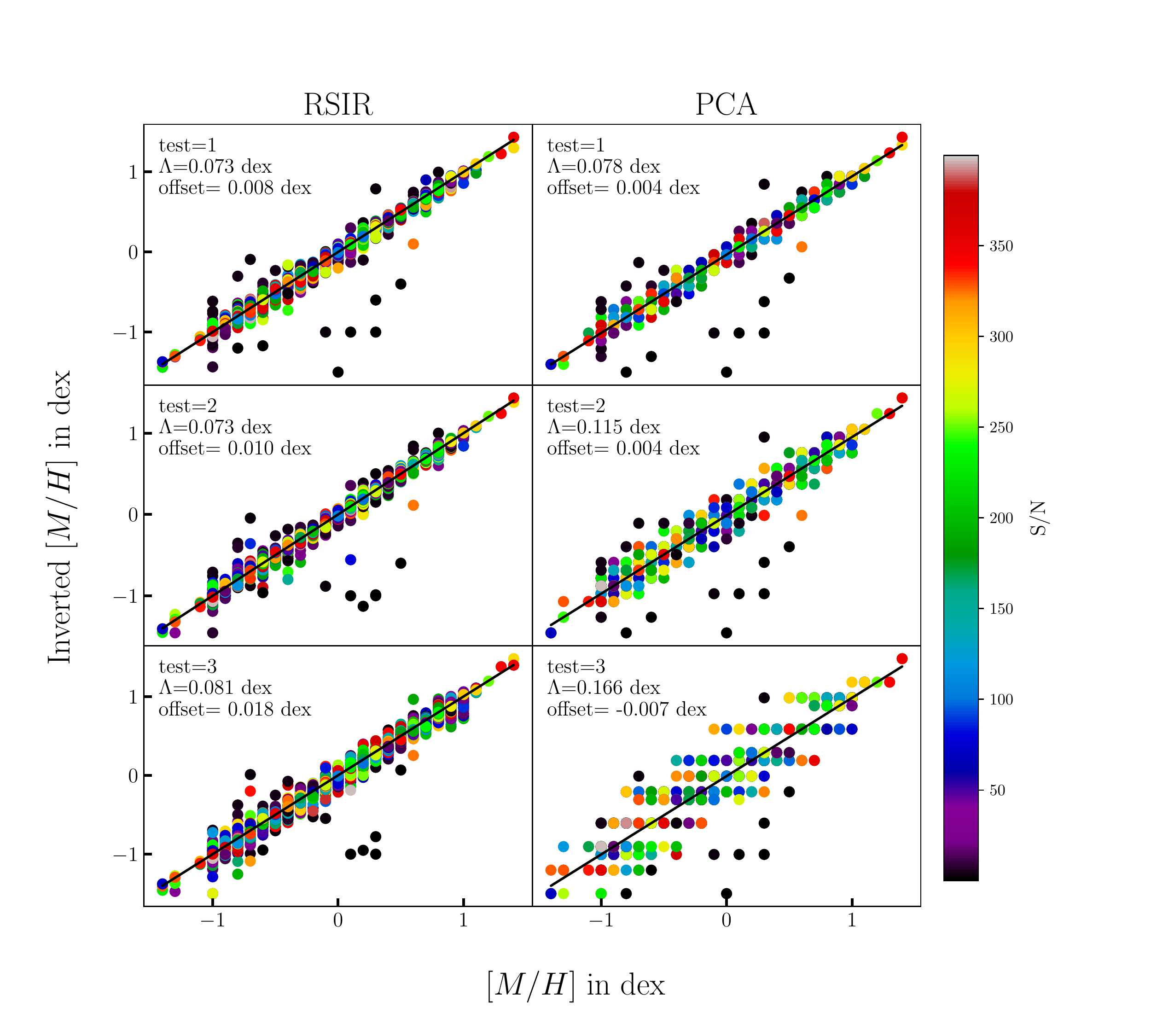}
    \caption{Results of the metallicity inversion for the synthetic A to K stars. }
  \label{fig meta_FGK}
\end{figure*}
\begin{figure*}
   \hspace*{-2cm} 	
   \includegraphics[scale=0.88]{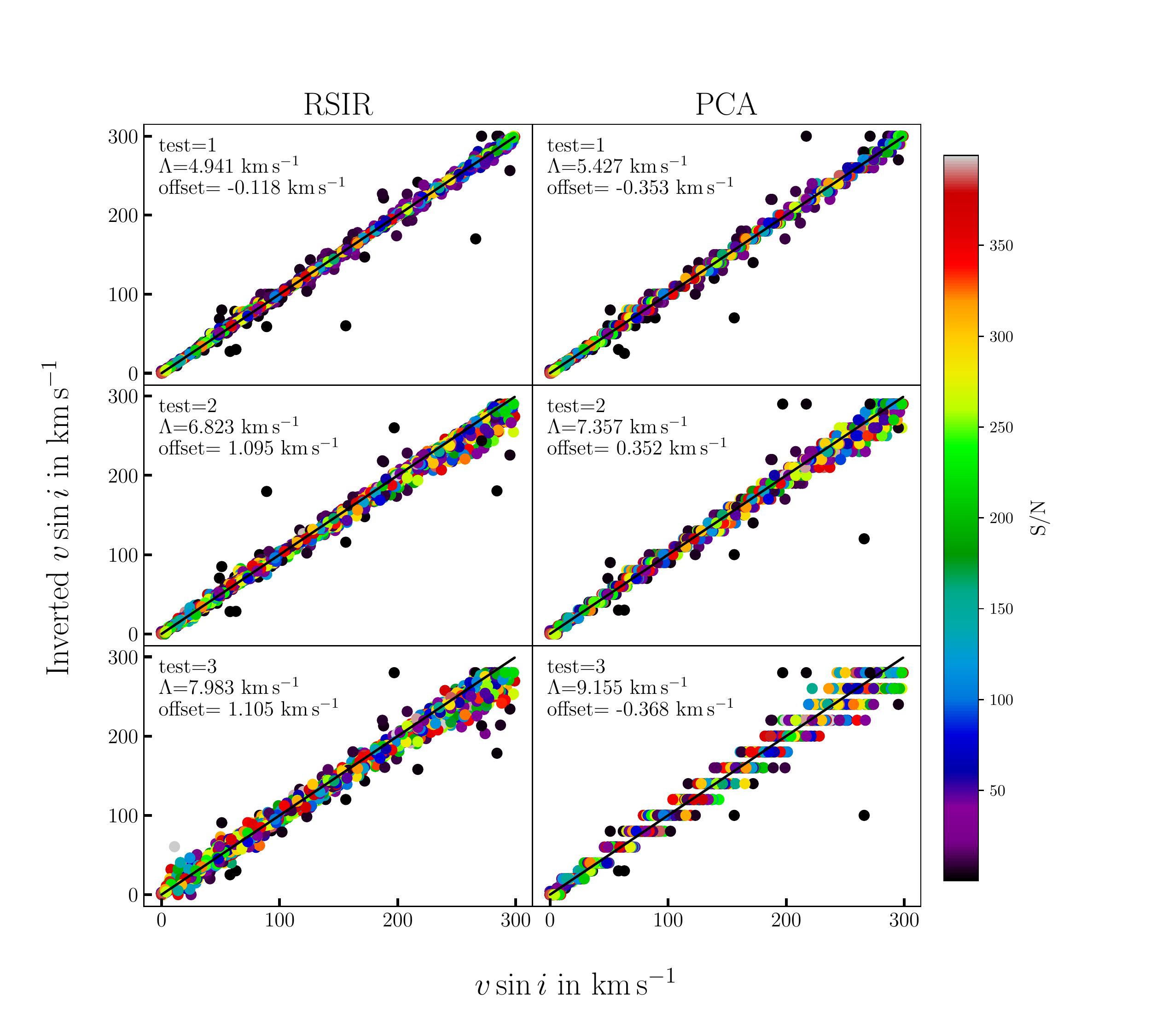}
    \caption{Results of projected equatorial velocity inversion for the synthetic A to K stars. }
  \label{fig vrot_FGK}
\end{figure*}
\end{center}
\end{document}